\documentclass[onecolumn,floatfix,superscriptaddress,a4paper,
               showpacs,showkeys,nofootinbib,preprint]{revtex4}
\usepackage{epsfig}
\usepackage{latexsym}
\usepackage{xspace}
\usepackage{hyperref}
\usepackage[latin2]{inputenc}
\usepackage{indentfirst}
\usepackage{enumerate}

\usepackage{amsmath}
\usepackage{amssymb}
\usepackage[english]{babel}
\usepackage{url}

\newcommand{\eq}[1]{\begin{align} #1 \end{align}}

\begin{document}

\title{Particle Number Fluctuations for \\ van der Waals Equation of State
}

\author{V. Vovchenko}
\affiliation{
Taras Shevchenko National University of Kiev, Kiev, Ukraine}
\affiliation{
Frankfurt Institute for Advanced Studies, Johann Wolfgang Goethe University, Frankfurt, Germany}
\affiliation{
GSI Helmholtzzentrum f\"ur Schwerionenforschung GmbH, Darmstadt, Germany}
\author{D. V. Anchishkin}
\affiliation{
Bogolyubov Institute for Theoretical Physics, Kiev, Ukraine}
\affiliation{
Taras Shevchenko National University of Kiev, Kiev, Ukraine}
\author{M. I. Gorenstein}
\affiliation{
Bogolyubov Institute for Theoretical Physics, Kiev, Ukraine}
\affiliation{
Frankfurt Institute for Advanced Studies, Johann Wolfgang Goethe University, Frankfurt, Germany}
\date{\today}

\begin{abstract}
The van der Waals (VDW) equation of state describes
a thermal equilibrium in system of particles, where both repulsive and attractive
interactions between them are included. This equation predicts an existence of the 1st order
liquid-gas phase transition and the critical point. The standard form of the VDW
equation is given by the pressure function in the canonical ensemble (CE) with
a fixed number of particles.
In the present paper the VDW equation is transformed to the grand canonical ensemble (GCE).
We argue that this procedure can be useful  for new physical applications.
Particularly, the fluctuations of number of particles, which are absent in the
CE, can be studied in the GCE.  For the VDW
equation of state  in the GCE the particle number fluctuations are calculated for the whole
phase diagram, both outside and
inside the liquid-gas mixed phase region. It is shown that
the scaled variance of these fluctuations remains finite within the mixed phase
and goes to infinity at the critical point.
The GCE formulation of the VDW equation of state can be also an important step
for its application to a statistical description of
hadronic systems, where numbers of different particle species are usually not conserved.

\end{abstract}

\pacs{12.40.-y, 12.40.Ee}

\keywords{}

\maketitle

\section{Introduction}
Statistical approach appears to be rather successful
in a description of hadron multiplicities
in high energy collisions. It leads to surprisingly
good agreement between the  results
of the hadron resonance gas
and the experimental data
(see, e.g.,
Refs.~\cite{Id-HRG-1,Id-HRG-2,Id-HRG-3,Id-HRG-4,Id-HRG-5,Id-HRG-6}).
The excluded volume (EV) model introduces the
effects of hadron repulsion
at short distances.
The EV procedure substitutes a system volume $V$
by the available volume
$V_{\rm av}=V- bN$~,
where $b$ is the particle
proper volume
parameter
and $N$ is the number of particles.
This substitution can be easily done in the
canonical ensemble (CE) formulation, where the
(conserved) particle number $N$ is an independent variable.
The ideal gas pressure $p_{\rm id}$ is then transformed to the EV pressure,
\eq{\label{ev-p-n}
p_{\rm id}(V,T,N)~=~\frac{NT}{V}~~\rightarrow~~p_{\rm ev}(V,N,T)~=~\frac{NT}{V~-~b\,N}~,
}
which is the function of volume $V$, temperature $T$, and number of particles $N$.
It was shown \cite{vdw-1,vdw-2} that
the substitution of $V$ by $V_{\rm av}$  leads to a shift of the
chemical potential $\mu$ and to a transcendental equation for the pressure
function in the grand canonical ensemble (GCE):
\eq{\label{ev-p-mu}
p_{\rm ev}(T,\mu)~=~p_{\rm id}[T,\mu-p_{\rm ev}(V,T,\mu)]~,
}
where $p_{\rm id}(T,\mu)$ is the pressure of the ideal gas calculated in
the GCE.
The EV model (\ref{ev-p-mu}) can be equivalently
formulated in terms of the mean-field (see Refs.~\cite{mf-1992,mf-1995,mf-2014}).
This makes it possible to incorporate other hadron interactions
within unified mean-field approach.
The
EV hadron equation of state is used for fitting the hadron multiplicities
(see, e.g., Refs.~\cite{ev-mult,ev-mult-1})
and for the hydrodynamic model calculations of nucleus-nucleus collisions
(see, e.g., Refs.~\cite{hyd-1,hyd-2,hyd-3}).

In the present paper
the equation of state
with both repulsive ($b>0$) and attractive ($a>0$) terms
(see, e.g., Refs.~\cite{greiner,LL}),
\eq{\label{vdw-p-n}
p(V,T,N)~=~\frac{NT}{V~-~b\,N}~-~a\,\frac{N^2}{V^2}~,
}
is considered.
The equation
of state (\ref{vdw-p-n}) was suggested by van der Waals (VDW) in 1873,
and
for his work he obtained the Nobel Prize in physics in 1910.
In the present paper the VDW equation (\ref{vdw-p-n}) will be
transformed to the GCE.
The GCE formulation will be then used to calculate the particle number
fluctuations in the VDW gas.

Normally, the GCE is an easiest one from the technical point of view.
Thus, it is used more frequently, and a typical problem in statistical
mechanics is to transform the GCE results to the CE ones.
However, the pressure functions of the EV (\ref{ev-p-n})
and VDW (\ref{vdw-p-n}) have their
simplest and transparent forms just in the CE.
A comparison of a simple analytical form (\ref{ev-p-n}) with a transcendental
equation (\ref{ev-p-mu}) elucidates this point.
On the other hand, the CE pressure
does not give a complete thermodynamical
description of the system. This is because the parameters
$V$, $T$, and $N$ are {\it not} the {\it natural}
variables for the pressure function.

Does one really need the GCE formulation
for the statistical description of hadrons?
There are at least three reasons for a positive answer to this question.

First,
the CE pressure, e.g., Eq.~(\ref{ev-p-n}) or Eq.~(\ref{vdw-p-n}),
does not give any information
concerning the particle mass $m$ and degeneracy factor $d$.
However,
the average energy of the system
depends on the value of $m$, e.g.,
the thermal part of energy per particle equals
to $m+3T/2$ at $m\gg T$, and to $3T $ at $m\ll T$.

Second, the number of hadrons of a given type is usually
not conserved. For example, the number of pions
can not be considered as an independent variable and is a function of
volume and temperature. The GCE formulation
of the VDW equation
can be therefore an important step
for inserting the repulsive and attractive VDW interactions
into the multi-component hadron gas.

Third, Eqs.~(\ref{ev-p-n}) and (\ref{vdw-p-n})
of the CE
give no answer about
the particle number fluctuations. Formally, the number of particles
does not fluctuate in the CE:
the total number of particles $N_0$ is constant in
the full volume $V_0$. However,  the number of particles $N$
starts to fluctuate if one considers
a sub-system with $V<V_0$.  If $V\ll V_0$ the $N$-fluctuations
should follow the GCE results. For the EV
model the particle number fluctuations
were calculated in Ref.~\cite{GHN}.

The paper is organized as follows.  In Sec.~\ref{CE-GCE} we discuss
how to transform the CE pressure to the GCE.
This procedure is illustrated by a derivation of the EV model
for $p_{\rm ev}(T,\mu)$ in the GCE.
In Sec.~\ref{sec-VDW} the VDW equation of state (\ref{vdw-p-n})
is transformed to the transcendental equation for particle number
density $n(T,\mu)$  in the GCE.
The particle number fluctuations and their behavior
in a vicinity of the critical point is then analyzed in Sec.~\ref{sec-fluc}.
A summary in Sec.~\ref{sec-sum} closes the article.

\section{From Canonical Ensemble to the Grand Canonical Ensemble}
\label{CE-GCE}
In the GCE the macroscopic states
of a thermal system are defined in terms of the {\it grand potential} $\Omega$
which is a function of 3 physical variables --
volume $V$, temperature $T$, and chemical potential $\mu$.
In a spatially homogeneous system
the grand potential $\Omega$ is straightforwardly connected to the system
pressure \cite{greiner}
\eq{\label{Omega}
\Omega(V,T,\mu)~=~-~p(T,\mu)\,V~.
}
Therefore,  pressure $p$ as a function of its {\it natural
variables}, temperature $T$ and chemical potential $\mu$, gives a
complete description of the equation of state. One calculates the
entropy density $s$, (conserved) number density $n$, and energy
density $\varepsilon$ as
\eq{
s(T,\mu)~=~\left(\frac{ \partial p}{\partial T} \right)_{\mu}~,
~~~~~~~~~~~~
n(T,\mu)~=~\left(\frac{ \partial p}{\partial \mu} \right)_{T}~,~~~~~~~~
\varepsilon(T,\mu)~=~Ts~+~\mu n~-~p~.
}

The thermodynamical potential in the CE
is the {\it free energy} $F(V,T,N)$ (or Helmholtz potential). The
function $F$ depends on its {\it natural variables}, volume $V$,
temperature $T$, and  number of particles $N$,
and it gives a complete thermodynamical description of the
system properties. Entropy $S$, pressure $p$, chemical potential
$\mu$, and total energy $E$ are calculated in the CE as
\cite{greiner}
\eq{
S(V,T,N)=- \left(\frac{\partial F}{\partial T}\right)_{V,N}~,~~~~
p(V,T,N)= -\left(\frac{\partial F}{\partial V}\right)_{T,N}~,~~~~
\mu(V,T,N) = \left(\frac{\partial F}{\partial N}\right)_{T,V}~,
}
and
\eq{
E(V,T,N)~=~F~+~TS~.
}

If one knows $p=p(V,T,N)$ in the CE it is not yet possible to find $S(V,T,N)$, $\mu(V,T,N)$,
and $E(V,T,N)$
in unique way.  For a complete thermodynamical description in
the CE one needs to reconstruct free energy $F$, thus, the
following equation should be solved:
\eq{\label{eqF}
\left(\frac{\partial F}{\partial V}\right)_{T,N}~ = ~-~ p(V,T,N)~.
}
The solution of Eq.~(\ref{eqF}) can be presented as
\eq{\label{solF}
F(V,T,N)~ = ~ F(V_0,T,N)~ - ~\int_{V_0}^{V} dV' p(V',T,N)~.
}
To find $F(V,T,N)$ one needs to know not only $p(V,T,N)$ but,
additionally, the free energy $F(V_0,T,N)$ as the function of $T$
and $N$ at some fixed value of $V=V_0$.
We assume that for fixed $T$ the role of particle interactions
becomes negligible at very small density $N/V_0\rightarrow 0$. Thus,
the ideal gas expression, $F(V_0,T,N)~\cong~ F_{\rm id}(V_0,T,N)$,
becomes valid at $V_0\rightarrow \infty$.
For relativistic ideal Boltzmann gas the free energy reads
\cite{greiner}
\eq{\label{Fid}
 F_{\rm id} (V,T,N)~ =~
 -~NT\,\left[1~+~\ln \frac{V\,\phi(T;d,m)}{N}\right]~ ,
}
where
\eq{\label{phi}
\phi(T;d,m)~ =~\frac{d}{2\pi^2}\int_0^{\infty} k^2 dk\, \exp(-\sqrt{k^2+m^2}/T)~=~
\frac{d\,m^2\, T}{2 \pi^2} \, K_2\left(\frac{m}{T}\right)~.
}
In Eqs.~(\ref{Fid}) and (\ref{phi}), $d$ is the degeneracy factor,
$m$ is the particle mass, and quantity $\phi$ has a physical meaning of the
GCE particle number density at zero chemical potential.

Let us start with an example of the EV volume equation (\ref{ev-p-n})
in the CE. The particle
proper volume
parameter in Eq.~(\ref{ev-p-n}) equals to $b = 4\cdot (4\pi r^3/3)$
with $r$ being the
corresponding hard
sphere radius of particle.
This equation describes the repulsion between particles and
can be rigorously obtained (in particular,
a factor of 4 in the
expression for $b$)
for a gas of the hard balls at low density (see, e.g., Ref. \cite{LL}).

From Eq.~(\ref{solF}) it follows
\eq{\label{eq:Fvdw}
&F(V,T,N)~ = ~
 -~NT\,\left[1~+~\ln \frac{V_0\,\phi(T;d,m)}{N}\right]~
+ ~\int_{V}^{V_0} dV' \frac{NT}{V'-bN}~\nonumber \\
&=~-~NT\,\left[1~+~\ln \frac{V_0\,\phi(T;d,m)}{N}\right] ~
+~NT \,\ln \frac{V_0-bN}{V-bN} ~=~
F_{\rm id}(V-bN,T,N)~,
}
where we have used that
$(V_0-bN)/V_0~\rightarrow ~1~$ at $V_0\rightarrow \infty $.
For the free energy (\ref{eq:Fvdw})
the chemical potential can be calculated as
\eq{\label{mu}
\mu~=~\left( \frac{\partial F}{\partial N}\right)_{V,T}~
=~-T~\ln \frac{(V-bN)\,\phi(T;d,m)}{N}~+~b\,
\frac{NT}{V-bN}~,
}

For particle number density $n=N/V$ one finds from Eq.~(\ref{mu}) for the ideal gas,
i.e., at $b=0$:
\eq{\label{nid}
\frac{N}{V}~=~n_{\rm id}(T,\mu)~=~\exp\left(\frac{\mu}{T}\right)\,\phi(T;d,m)~.
}
At $b>0$ one can easily find from Eq.~(\ref{mu}):
\eq{\label{nev}
\frac{N}{V}~\equiv~n_{\rm ev}(T,\mu)~
=~\frac{n_{\rm id}(T,\mu^*)}{1~+~b\,n_{\rm id}(T,\mu^*)}~,
~~~~\mu^*~=~
\mu~-~b\,\frac{n_{\rm ev}(T,\mu)\,T}{1\,-\,b\,n_{\rm ev}(T,\mu)}~,
}
or equivalently,
\eq{\label{pev}
p_{\rm ev}(T,\mu)~=~
p_{\rm id}(T,\mu^*)~,~~~~
\mu^*~=~\mu~-~b\,\frac{n_{\rm ev}(T,\mu)\,T}{1\,-\,b\,n_{\rm ev}(T,\mu)}
=\mu~-~b\,p_{\rm ev}(T,\mu)~.
}
Therefore,
to solve the excluded-volume problem in case of the Boltzmann statistics
one has to obtain solution of transcendental equation (\ref{nev})
with respect to particle density  $n_{\rm ev}(T,\mu)$
or of transcendental equation (\ref{pev}) with respect to pressure
$p_{\rm ev}(T,\mu)$ for every point in $(\mu,\,T)$-plane, i.e., in the GCE.
These equations coincide with those suggested in Refs.~\cite{mf-1992, mf-1995, mf-2014}
and in Refs.~\cite{vdw-1,vdw-2}, respectively.

\section{Van der Waals equation of state in the GCE}
\label{sec-VDW}
\subsection{GCE formulation}
The VDW  equation for the system pressure is defined in the CE in the form of
Eq.~(\ref{vdw-p-n}).
Using Eq.~(\ref{solF}) one obtains
\eq{\label{F-vdw}
F(V,T,N)~ &= ~F_{\rm id}(V_0,T,N)
~ + ~\int_{V}^{V_0} dV'
\left[\frac{NT}{V'-bN}~-~a\frac{N^2}{V'^2}\right]\nonumber \\
&\cong ~ F_{\rm id}(V-bN,T,N)~-~a\,\frac{N^2}{V}~.
}
Chemical potential then reads
\eq{\label{muvdw}
\mu~=~\left( \frac{\partial F}{\partial N}\right)_{V,T}~
=~-T~\ln \frac{(V-bN)\,\phi(T;d,m)}{N}~+~b\,
\frac{NT}{V-bN}~-~2a\,\frac{N}{V}~,
}
and from Eq.~(\ref{muvdw}) one obtains
\eq{\label{nvdw}
\frac{N}{V}~\equiv~n(T,\mu)~=~\frac{n_{\rm id}(T,\mu^*)}{1~+~b\,n_{\rm id}(T,\mu^*)}~,
~~~~~~\mu^*~=~
\mu~-~b\,\frac{n\,T}{1-b\,n}~+~2a\,n~.
}
The function $n(T,\mu)$  which is the solution of transcendental
equation (\ref{nvdw}) is reduced to $n_{\rm ev}(T,\mu)$
(\ref{nev}), if $a=0$, and to $n_{\rm id}(T,\mu)$ (\ref{nid}), if
$a=b=0$. At fixed $T$ and $\mu$, a presence of parameter $b>0$, which describes
the repulsion between particles, leads
to a suppression of particle number density $n(T,\mu)$, while attractive interactions
described by $a>0$ lead to its enhancement.

The energy density of the VDW gas can be calculated within the GCE
from the pressure function, $p(T,\mu)$,
\eq{\label{e-vdw-gce}
\varepsilon(T,\mu)&=T\,\left(\frac{\partial p}{\partial T}\right)_{T} +
n\,\left(\frac{\partial p}{\partial \mu}\right)_{\mu}~-~p=
\frac{\varepsilon_{\rm id}(T,\mu^*)}{1~+~b\,n_{\rm id}(T,\mu^*)}- a\, n^2~\nonumber \\
&=~\Big[\overline{\epsilon}_{\rm id}(T;m)~-~a\,n\Big]\,n~,
}
where $\widetilde{\mu}$ is defined in Eq.~(\ref{nvdw}) and
\eq{\label{epsilon}
\overline{\epsilon}_{\rm id}(T;m)~
=~\frac{\int_0^{\infty}k^2dk\,\sqrt{m^2+k^2}\,\exp\left(-\,\sqrt{m^2+k^2}/T\right)}
{\int_0^{\infty}k^2dk\,\exp\left(-\,\sqrt{m^2+k^2}/T\right)}~=~
3\,T + m \, \frac{K_1(m/T)}{K_2(m/T)}~
}
is the average energy per particle in the ideal gas.
A comparison of the right hand side of Eq.~(\ref{e-vdw-gce})
with the corresponding ideal gas expression, $\varepsilon_{\rm id}=
\overline{\epsilon}_{\rm id}(T;m)\,n_{\rm id}$,
demonstrates the role of
the $a$ and $b$ parameters for the system energy density.
Both $a$ and $b$ parameters influence the particle number density
and transform  $n_{\rm id}$ into $n$:
$a>0$ leads to an enhancement of particle number density and $b>0$
to its suppression.   Parameter $b$ does not contribute
however to the average energy per particle, whereas
the mean field $-a\,n$ created due to parameter $a$ is added
to the average energy per particle and changes it  from $\overline{\epsilon}_{\rm id}$
to $\overline{\epsilon}=\overline{\epsilon}_{\rm id}-a\,n$.

It is also instructive to consider another derivation of Eq.~(\ref{nvdw}) which
based on the procedure  used in Ref.~\cite{kost}.
Introducing particle number density, $n=N/V$, the VDW pressure
function (\ref{vdw-p-n}) can be written as
\eq{\label{pnT}
 p(T,n)~=~\frac{n\,T}{1-b\,n}~ -a\,n^2~.
 }
For the particle number density in the GCE the following differential equation
can be obtained
\eq{\label{nTmu}  n(T,\mu)~&\equiv ~\left(\frac{\partial
p(T,\mu)}{\partial \mu}\right)_T~ =~\left(\frac{\partial
p(T,n)}{\partial n}\right)_T
~\left(\frac{\partial n(T,\mu)}{\partial \mu}\right)_T~\nonumber \\
& =~
\left[\frac{T}{(1-b\,n)^2}~-~2a\,n\right]\,\left(\frac{\partial
n(T,\mu)}{\partial \mu}\right)_T~.
}
A general solution of Eq.~(\ref{nTmu})  can be rewritten as
\eq{\label{nTmu-1}
%
\mu=\int dn\left[ \frac{T}{n\,(1-b n)^2}-2a\right]=
T\,\ln\left[\frac{n}{\phi(T;d,m)\,(1-b n)}\right]+ \frac{bT\, n}{1-b\, n}-2a\, n~.
}
The temperature dependent integration constant in the solution (\ref{nTmu-1}) is defined in
the form which guarantees the ideal gas solution (\ref{nid}) at $b=a=0$.
One can easily check that functions  $n(T,\mu)$ defined by Eq.~(\ref{nTmu-1})
and Eq.~(\ref{nvdw}) are identical.

\subsection{Phase Diagram
for the VdW Gas}\label{phase-diag}

The VDW equation of state contains a 1$^{\rm st}$ order liquid-gas phase
transition and has a critical point.
The thermodynamical quantities at the critical point are equal to \cite{LL}:
\eq{\label{crit}
 T_c = \frac{8a}{27b}~,~~~~~ n_c =
\frac{1}{3b}~,~~~~~ p_c = \frac{a}{27b^2}~.
}
\begin{figure}[t]
\begin{minipage}{.49\textwidth}
\centering
\includegraphics[width=\textwidth]{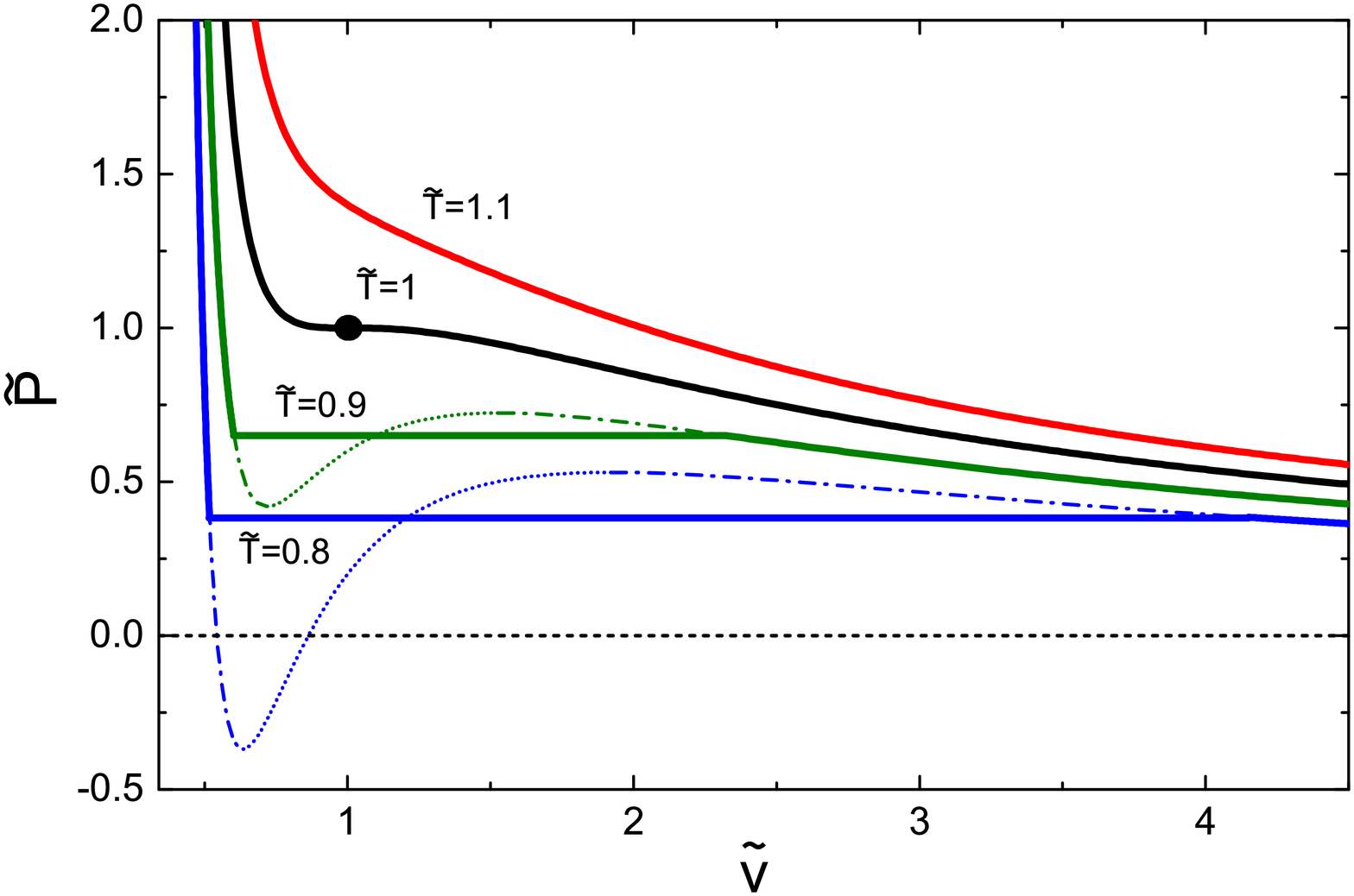}
\end{minipage}
\begin{minipage}{.49\textwidth}
\centering
\includegraphics[width=\textwidth]{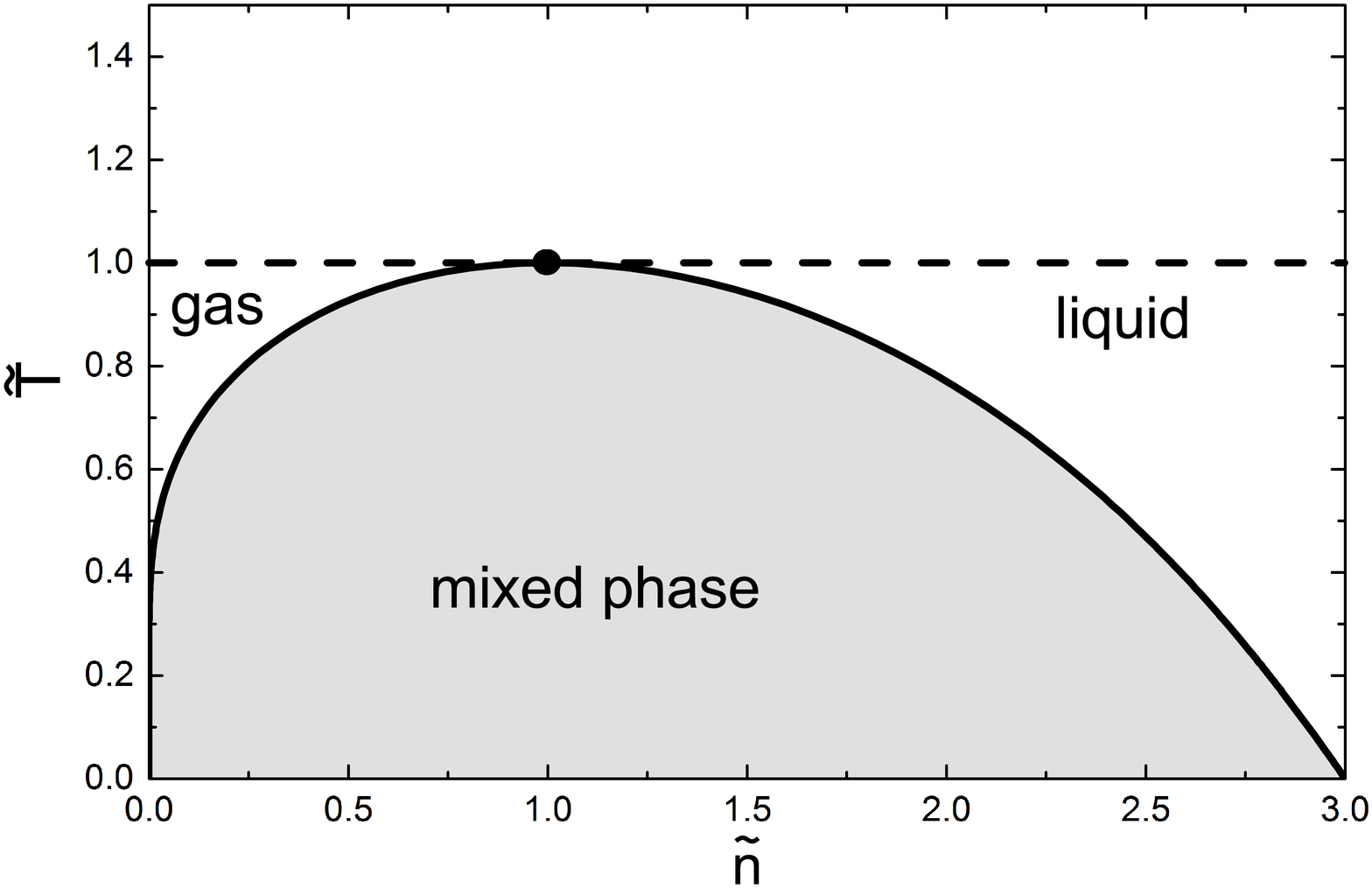}
\end{minipage}
\caption{(Color online) (a): The dimensionless form of the VDW
isotherms for pressure, $\widetilde{p}$,
versus the volume per particle
$\widetilde{v}=\widetilde{n}^{-1}$.
The dashed-dotted lines
present the metastable parts of the VDW
isotherms at $\widetilde{T}<1$, whereas
the dotted lines correspond to unstable parts.  The full circle
on the $\widetilde{T}=1$ isotherm corresponds to the critical point.
(b): The phase diagram for the VDW equation of state
on the $(\widetilde{n},\widetilde{T})$ plane.
The phase coexistence region
resulting from the Maxwell construction is depicted by grey shaded
area.
 } \label{fig:dim-isotherms}
\end{figure}

The VDW equation (\ref{vdw-p-n}) can be then rewritten in the following
dimensionless (reduced) form:
\begin{equation}
\widetilde{p}\ =\ \frac{8\,\widetilde{T}\, \widetilde{n}}{3 -
\widetilde{n}}\, -\, 3\,\widetilde{n}^2 \,, \label{vdw-dim}
\end{equation}
where $\widetilde{n} = n/n_c$, $\widetilde{p} = p/p_c$, and
$\widetilde{T} = T/T_c$.
In the dimensionless presentation (\ref{vdw-dim}) the VDW equation
has a  universal form independent of the values of $a$ and $b$, and
the critical point
(\ref{crit}) is transformed to
$\widetilde{T}_c \, =\, \widetilde{p}_c\, =\,
\widetilde{n}_c\, =\, 1 \,.$

The dimensionless VDW isotherms
are presented in
Fig.~\ref{fig:dim-isotherms} (a) as functions of $\widetilde{v}\equiv \widetilde{n}^{-1}$ .
To describe the phase coexistence
region below the critical temperature the VDW isotherms
should be corrected by the well-known Maxwell construction of {\it equal areas}.
These corrected parts of the VDW isotherms are shown by the solid horizontal lines.
Figure \ref{fig:dim-isotherms} (b)
depicts the liquid-gas  coexistence region on
the $(\widetilde{n},\widetilde{T})$ plane.
The pure gaseous phase exists at all $\widetilde{T}<1$. However,
it can be hardly seen at $\widetilde{T}\ll 1$ in Fig.~\ref{fig:dim-isotherms} (b). This is because
of very small admittable  values of $\widetilde{n}$ inside the pure gaseous phase
at $\widetilde{T}\rightarrow 0$, e.g.,
the largest value of the particle number density in the gaseous phase
at $\widetilde{T}=0.25$  is $\widetilde{n}\cong 5 \cdot 10^{-5}$.
At any value of $\widetilde{T}<1$, the particle number density in the mixed phase is given by
\eq{\label{n-mix}
n~=~\xi n_g~+~(1-\xi) n_l~,
}
where $\xi$ and $1-\xi$ are the volume fractions of the gaseous
and liquid components, respectively.
The values of $n_g$ and $n_l$ in Eq.~(\ref{n-mix})
are the particle densities of, respectively, the gaseous and liquid phases
at the corresponding boundaries with the mixed phase. For example, at $\widetilde{T}=0.25$  one finds $\widetilde{n}_g\cong 5 \cdot 10^{-5}$
and $\widetilde{n}_l\cong 2.75$. For $\xi=0.95$ Eq.~(\ref{n-mix}) then
gives $\widetilde{n}\cong 0.95\cdot  5 \cdot 10^{-5}+ 0.05 \cdot 2.75 \cong 0.14$. Therefore, at $\widetilde{n}=0.14$
and $\widetilde{T}=0.25$ the gaseous component occupies 95\% of the total volume of the mixed state
but gives a negligible contribution to the total number of particles.
Note also that $\widetilde{n}=3$ is an upper
bound for the dimensionless particle density.

Let us consider the Maxwell rule in more details.
At each $\widetilde{T}<1$ the VDW loops appear in the isotherm
$\widetilde{p}(\widetilde{v})$. They are shown by the dashed lines
in Fig.~\ref{fig:dim-isotherms} (a). Each loop consists from {\it metastable}
part, where $\partial \widetilde{p}/ \partial\widetilde{v}<0$,
and {\it unstable} part, where $\partial \widetilde{p}/ \partial\widetilde{v}>0$.
They are shown in Fig.~\ref{fig:dim-isotherms} (a) by the dashed-dotted and dotted lines, respectively.
According to the Maxwell rule, the loop between  values of $\widetilde{v}_l$ and $\widetilde{v}_g$
is substituted
by the constant value $\widetilde{p}_0$ (solid horizontal line in Fig.~\ref{fig:dim-isotherms} (a)),
and the following condition should be
fulfilled:
\eq{\label{Max}
\int_{\widetilde{v}_l}^{\widetilde{v}_g} d\widetilde{v} \, \widetilde{p}~=~\widetilde{p}_0\,
\left(\widetilde{v}_g~-~\widetilde{v}_l\right)~.
}

It can be proven that the GCE formulation forbids the Maxwell loops
and it leads automatically to the Maxwell construction.
Let us assume that the system can be divided into two parts -- {\it gas} and {\it liquid} --
with
\eq{\label{VgVl}
N_g~+~N_l~=~N~,~~~~~V_g~+~V_l~=~V~,~~~ {\rm and  } ~~\frac{N_g}{V_g}~\neq ~\frac{N_l}{V_l}~.
}
The free energy of the system can be then presented as
\eq{\label{F-gl}
F(V,T,N)~=~F(V_g,T,N_g)~+~F(V_l,T,N_l)~,
}
where the surface free energy is neglected in the thermodynamic limit $V\rightarrow \infty$.
An equilibrium state corresponds to a minimum of the free energy.
This leads to
\eq{\label{F-min}
\left(\frac{\partial F}{\partial V_g}\right)_{T,N_g}~=0~,~~~~~~
\left(\frac{\partial F}{\partial N_g}\right)_{T,V_g}~=0~.
}
Relations in Eq.~(\ref{F-min}) are equivalent, respectively, to
\eq{\label{mix-eq}
p(T,V_g,N_g)~=~p(T,V_l,N_l)~,~~~~~\mu(T,V_g,N_g)~=~\mu(T,V_l,N_l)~.
}
The first condition in Eq.~(\ref{mix-eq}) means that the Maxwell loop
is transformed to the mixed phase region with equal pressures. And the
second condition in Eq.~(\ref{mix-eq}) leads in a unique way to the Maxwell rule
of equal areas.
The equilibrium conditions of phase coexistence are then fulfilled:
both phases have the same temperature, pressure and chemical
potential in the coexistence region. These conditions are
known as the Gibbs rules for a
1$^{\rm st}$ order phase transition \cite{LL}.

\section{Particle Number Fluctuations in the VdW Gas}\label{sec-fluc}

\subsection{Particle Number Fluctuations in the Pure Phases}
The variance of the total particle number in the GCE can be
calculated as
\eq{\label{Var} Var[N]~\equiv~\langle N^2 \rangle~-~\langle
N\rangle^2~ =~ T \left(\frac{\partial{\langle N \rangle}}{\partial
\mu}\right)_{T,V} = T \, V \, \left(\frac{\partial{n}}{\partial
\mu}\right)_{T},
}
where symbol $\langle ...\rangle$ denotes the GCE averaging, and
$n(T,\mu)$ is the particle number density in the GCE. The scaled
variance for the particle number fluctuations is then
\eq{\label{omega}
\omega[N]~\equiv~\frac{Var[N]}{\langle
N\rangle}~ =~ \frac{T}{n} \, \left(\frac{\partial{n}}{\partial
\mu}\right)_{T}~.
}
It will be convenient to rewrite Eq.~\eqref{nvdw} as
\eq{\label{nvdw2} n~ =~ (1-bn) \, n_{\rm id}(T, \mu) \,
\exp\left(-\frac{bn}{1-bn}\right) \,
\exp\left(\frac{2an}{T}\right)~. }
Taking the derivative of Eq.~\eqref{nvdw2} with respect to $\mu$
one obtains
\eq{
\frac{\partial n}{\partial \mu} = \frac{n}{T}
\left[\frac{1}{(1-bn)^2} - \frac{2an}{T} \right]^{-1}.
}
Therefore, the scaled variance (\ref{omega}) equals to
\eq{\label{omega-1}
\omega[N]~ =~ \left[\frac{1}{(1-bn)^2} - \frac{2an}{T}
\right]^{-1}~.
}
It is clearly seen from Eq.~(\ref{omega-1}) that in the VDW
gas the repulsive interactions suppress the particle number
fluctuations, whereas the attractive interactions lead to their
enhancement. Note that for $a=0$ the value of $\omega[N]$
(\ref{omega-1}) coincide with the result  for the EV
model obtained in Ref.~\cite{GHN}.

\begin{figure}[t]
\begin{minipage}{.8\textwidth}
\centering
\includegraphics[width=\textwidth]{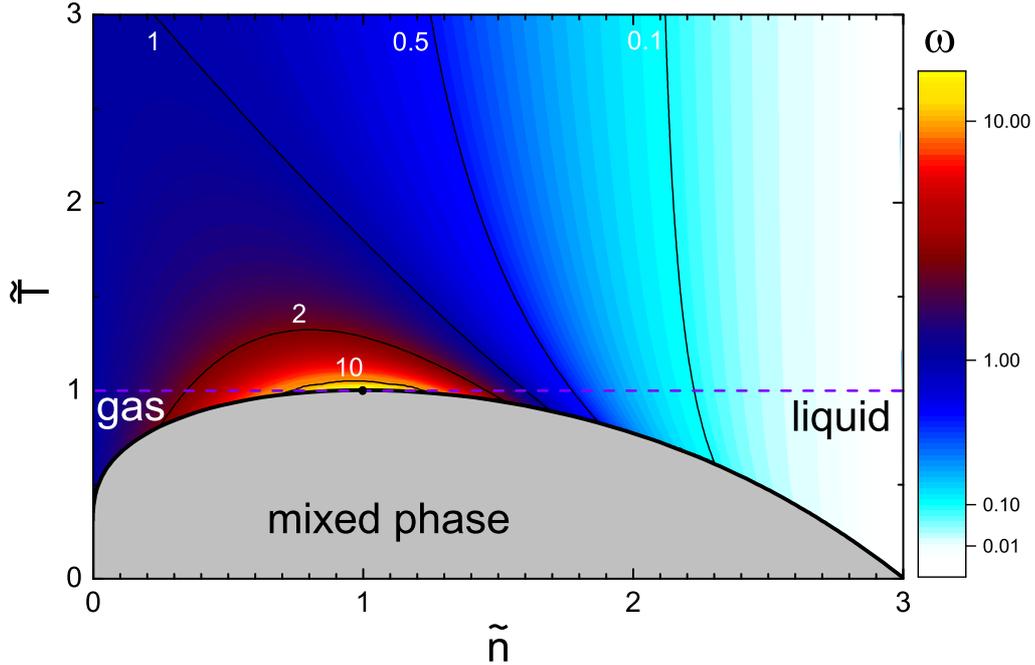}
\end{minipage}
\caption{(Color online)
The lines of constant values of the scaled variance $\omega[N]$ are shown
on the $(\widetilde{n},\widetilde{T})$ phase diagram,
outside of the mixed phase region.
} \label{fig-fluct}
\end{figure}

The scaled variance (\ref{omega-1}) expressed in terms $\widetilde{n}$ and
$\widetilde{T}$ equals to
\eq{\label{omega-2}
\omega[N]~ =~ \frac{1}{9}\,\left[\frac{1}{(3~-~\widetilde{n})^2}
~-~ \frac{\widetilde{n}}{4\,\widetilde{T}} \right]^{-1}~.
}
In Fig.~\ref{fig-fluct} the lines of constant values of $\omega[N]$ are shown
on the $(\widetilde{n},\widetilde{T})$ phase diagram outside of the mixed phase region.
At any fixed value of $\widetilde{T}$, the particle number fluctuations (\ref{omega-2})
approach to those of the ideal gas, i.e.  $\omega[N]\cong 1$,
at $\widetilde{n}\rightarrow 0$,
and become small, $\omega[N]\ll 1$, at $\widetilde{n}\rightarrow 3$.
As it should be, the scaled variance (\ref{omega-2}) is always positive for all
possible values of $\widetilde{n}$ and $\widetilde{T}$ outside of the mixed phase region.
At $\widetilde{T}\rightarrow 0$ this is insured by simultaneous limiting behavior
of $\widetilde{n}\rightarrow 0$ in the pure gaseous phase and $\widetilde{n}\rightarrow 3$
in the pure liquid phase.

At the critical point ($\widetilde{T}=\widetilde{n}=1$)
the scaled variance of the particle number fluctuations diverges in the  GCE.
To study the behavior of $\omega[N]$ in a vicinity of the critical point
we introduce the quantities $\tau =\widetilde{T}-1 \ll 1$
and $\rho=\widetilde{n}-1\ll 1$.
Expanding (\ref{omega-2}) at small $\tau$
and $\rho$ and keeping only the lowest orders
of them one finds
\eq{\label{omega-c}
\omega[N]~\cong ~\frac{4}{9}\,\left[ \tau~+~ \frac{3}{4}\rho^2~+~ \tau\,\rho\right]^{-1}~.
}
In particular,
\eq{\label{omega-c}
\omega[N]~\cong~ \frac{4}{9}\,\tau^{-1}~~ {\rm at}~~\rho=0~,
~~~~~~ {\rm and}~~\omega[N]~\cong~ \frac{16}{27}\,\rho^{-2}~~
{\rm at}~~ \tau=0~.
}
Note that thermodynamical parameters
$\widetilde{T}$ and $\widetilde{n}$ correspond to
points outside the mixed phase region. Thus,
in Eq.~(\ref{omega-c}), parameter $\tau$ is positive,
while $\rho$ can be both positive and negative.

\begin{figure}[t]
\begin{minipage}{.8\textwidth}
\centering
\includegraphics[width=\textwidth]{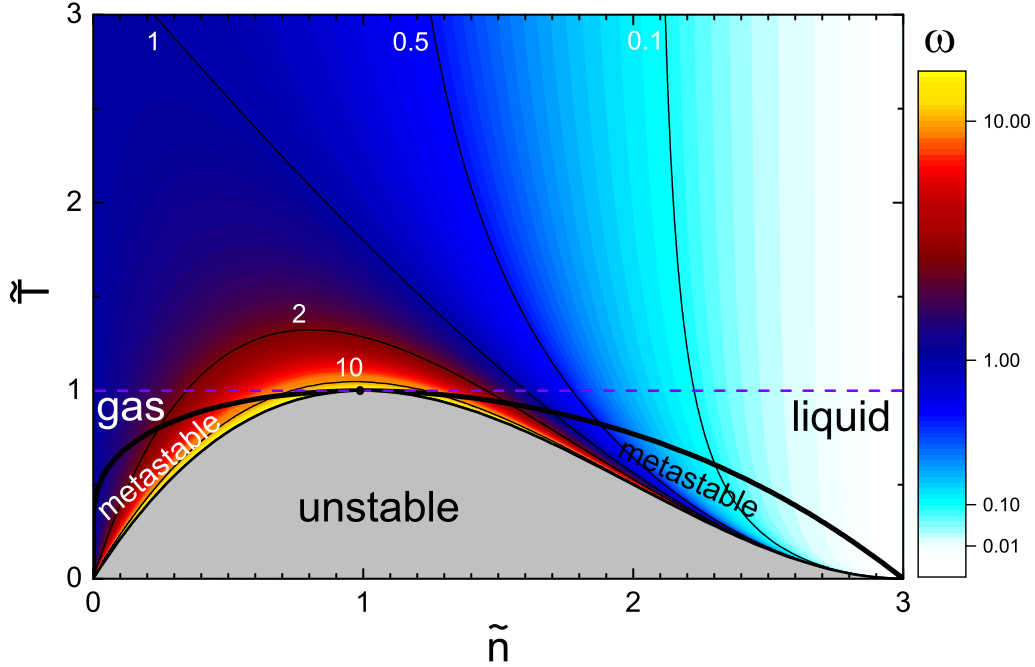}
\end{minipage}
\caption{
(Color online)
The lines of constant values of the scaled variance $\omega[N]$ are shown
on the $(\widetilde{n},\widetilde{T})$ phase diagram for
both stable and metastable pure phases.
The boundary between stable and metastable phases is depicted by the thick
black line, and the unstable region is depicted by the gray area.
} \label{fig-fluct-MS}
\end{figure}

As was mentioned above, the VDW equation of state permits the existence
of metastable phases of super-heated liquid and
super-cooled gas. These states are depicted by the
dash-dotted lines on the VDW isotherms in Fig.~\ref{fig:dim-isotherms} (a).
In metastable phases the system is assumed to be uniform
and, therefore, one can use Eq.~(\ref{omega-2})
to calculate particle number fluctuations in
these phases. In Fig.~\ref{fig-fluct-MS}  the lines of constant values of $\omega[N]$ are shown
on the $(\widetilde{n},\widetilde{T})$ phase diagram for
both stable and metastable pure phases, while the unstable region is depicted by the gray area.
It is seen that the scaled variance remains finite,
and diverges only at the boundary between the metastable and unstable regions.
We recall that at this boundary $\partial \widetilde{p} / \partial \widetilde{n} = 0$,
where $\widetilde{p}$ is the dimensionless CE pressure~\eqref{vdw-dim}. One can easily show
using Eqs.~\eqref{vdw-dim} and \eqref{omega-2} that $\omega[N] \to \infty$
when $\partial \widetilde{p} / \partial \widetilde{n} = 0$. Note that metastable
regions of the equation of state can be reached within fast non-equilibrium
processes, whereas the unstable region
is physically forbidden. Note that the thermodynamical relations are
not fulfilled in the unstable region, e.g., nonphysical behavior with
$\omega[N] < 0$ is found in this region.

\subsection{Particle Number Fluctuations in the Mixed Phase}
Let us consider a point $(n,T)$ inside the mixed phase.
The volume $V$ is then divided into the two parts,
$V_g=\xi V$ occupied by the gaseous phase with particle number density $n_g$
and $V_l=(1-\xi)V$ occupied by the liquid phase with
particle number density $n_l$. The total number of particles $N$
equals to $N_g+N_l$. The average value and scaled variance
for the $N$-distribution are
\eq{\label{Ntot}
& \langle N\rangle~=~\langle N_g\rangle~+~\langle N_l\rangle
~=~V\,[\xi n_g ~+~(1-\xi) n_l]~,\\
&\omega[N]~=~\frac{\langle N^2\rangle -\langle N\rangle^2}{\langle N\rangle}~=~
\frac{\langle N_g^2\rangle -\langle N_g\rangle^2}{\langle N\rangle}~ +~
\frac{\langle N_l^2\rangle -\langle N_l\rangle^2}{\langle N\rangle}~
+~2\frac{\left[\langle N_g N_l \rangle
-\langle N_g\rangle \langle N_l\rangle\right]}{\langle N\rangle}~.~
\label{omega-N}
}

In calculating the variations of $N_g$ and $N_l$ distributions inside the
the mixed phase, one should take into account
the fluctuations of the corresponding volumes $V_g=\xi V$ and $V_l=(1-\xi)V$.
In fact, this is the fluctuations of the parameter $\xi$.
In the thermodynamic limit $V\rightarrow \infty$, one finds:
\eq{
& \frac{\langle N_g^2\rangle - \langle N_g \rangle^2}
{\langle N\rangle}~=~\frac{\langle N_g\rangle}
{\langle N\rangle}\,\omega_{\xi}[N_g]~+~
\frac{n_g^2}{n} \, V \left[\langle \xi^2\rangle
-\langle \xi\rangle^2\right]~, \label{omega-Ng}\\
& \frac{\langle N_l^2\rangle - \langle N_l \rangle^2}
{\langle N\rangle}~=~\frac{\langle N_g\rangle}
{\langle N\rangle}\,\omega_{\xi}[N_l]~+~
\frac{n_l^2}{n} \, V \left[\langle \xi^2\rangle
-\langle \xi\rangle^2\right] ~,\label{omega-Nl}\\
& \frac{\langle N_g N_l\rangle
- \langle N_g \rangle\langle N_l\rangle}{\langle N\rangle}~=~
-~2\,\frac{n_g n_l}{n} \, V \left[\langle \xi^2\rangle
-\langle \xi\rangle^2\right]~,\label{NgNl}
}
where $\omega_{\xi}[N_g]$ and $\omega_{\xi}[N_l]$
in Eqs.~(\ref{omega-Ng}) and (\ref{omega-Nl}) correspond
to the fixed value of $\xi$, and can be calculated using
Eqs.~(\ref{Var}) and (\ref{omega}).
One finds
\eq{\label{Var-mix}
\omega[N]~& =~
\frac{\xi_0 n_g}{n}\,\left[\frac{1}{(1-bn_g)^2} - \frac{2an_g}{T}
\right]^{-1}~+~~\frac{(1-\xi_0) n_l}{n}\,\left[\frac{1}{(1-bn_l)^2} - \frac{2an_l}{T}
\right]^{-1}~ \nonumber \\
&+~\frac{(n_g-n_l)^2\,V}{n} \left[\langle \xi^2\rangle -\langle \xi\rangle^2\right]~,
}
where the equilibrium value $\langle \xi\rangle\equiv \xi_0$ is defined as
\eq{\label{xi-eq}
\xi_0~=~\frac{n_l~-~n}{n_l~-~n_g}~.
}

To calculate the variance $\langle \xi^2\rangle -\langle \xi\rangle^2$ in Eq.~(\ref{Var-mix})
we can return to the CE formulation and use Eq.~(\ref{F-gl}) rewriting it as the following
\eq{\label{F-xi}
F(V,T,N;\xi)~=~F\left(\xi\,V,T,N_g\right)~+~F\left[(1-\xi)\,V,T,N_l\right]~.
}
Note that in the  limit $V\rightarrow \infty$
assumed in our study the particle number densities $n_g$ and $n_l$
in the GCE are identical to their CE values as a
consequence of thermodynamical equivalence.
  The probability distribution  $W(\xi)$  is proportional to
$\exp[-\,F(V,T,N;\xi)/T]$. The function $F$ is given by Eq.~(\ref{F-xi}) and
s a function of $\xi$ it can be presented as a power series expansion
over $\xi-\xi_0$ in a vicinity of the equilibrium value $\xi_0$.
This gives the $\xi$ normalized probability distribution ($C$ is a normalization factor)
in the form
\eq{\label{xi-distr}
W(\xi)~=~C~\exp\left[-\,\frac{1}{2T}\, \left(\frac{\partial ^2F}{\partial ^2 \xi}\right)_{\xi=\xi_0}\,
(\xi-\xi_0)^2\,\right]~\equiv~
C\,\exp\left[\,-\,\frac{(\xi-\xi_0)^2}{2\sigma^2}\right]~,
}
where
\eq{\label{sigma}
 \sigma^2~& =~T\,\left(\frac{\partial^2 F}{\partial ^2\xi}\right)^{-1}_{\xi=\xi_0}~=~
-\,\frac{T}{V^2}\,\left[\left(\frac{\partial p_g}{\partial V_g}\right)~+~
\left(\frac{\partial p_l}{\partial V_l}\right)\right]^{-1}~\nonumber \\
& =~\frac{T}{V}\,\left[\frac{n_g T}{\xi_0\,(1-b n_g)^2}-\frac{2a n_g^2}{\xi_0}
+ \frac{n_l T}{(1-\xi_0)(1-b n_l)^2}-\frac{2a n_l^2}{1-\xi_0} \right]^{-1}
}
\begin{figure}[t]
\begin{minipage}{.8\textwidth}
\centering
\includegraphics[width=\textwidth]{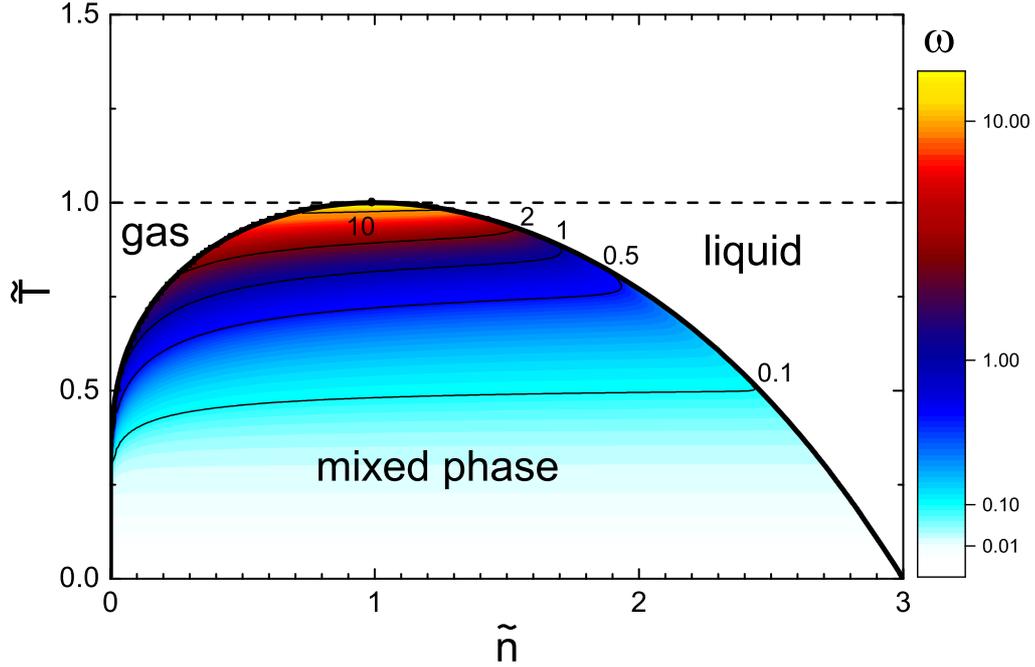}
\end{minipage}
\caption{(Color online) The lines of constant values of $\omega[N]$ are shown
on the $(\widetilde{n},\widetilde{T})$ phase diagram inside the mixed phase region.
 } \label{fig-fluct-mp}
\end{figure}
Using the $\xi$-distribution (\ref{xi-distr}) one finds
\eq{\label{var-xi}
\langle \xi^2\rangle~-~\langle \xi\rangle^2~=~\sigma^2~
}
and, finally,
\eq{\label{Var-mix1}
\omega[N]~& =~
\frac{\xi_0 n_g}{n}\,\left[\frac{1}{(1-bn_g)^2} - \frac{2an_g}{T}
\right]^{-1}~+~~\frac{(1-\xi_0) n_l}{n}\,\left[\frac{1}{(1-bn_l)^2} - \frac{2an_l}{T}
\right]^{-1}~ \nonumber \\
&+~\frac{(n_g-n_l)^2}{n}\, \left[\frac{n_g }{\xi_0(1-b n_g)^2}-\frac{2a n_g^2}{\xi_0\,T}
+ \frac{n_l }{(1-\xi_0)(1-b n_l)^2}-
\frac{2a n_l^2}{(1-\xi_0)T}\right]^{-1}~ .
}
Introducing dimensionless variables (\ref{vdw-dim}) one can rewrite Eq.~(\ref{Var-mix1})
as
\eq{\label{omega-mix}
\omega[N]~& =~
\frac{\xi_0 \widetilde{n_g}}{9\widetilde{n}}\,\left[\frac{1}{(3-\widetilde{n_g})^2} -
\frac{\widetilde{n_g}}{4\widetilde{T}}
\right]^{-1}~+~ \frac{(1-\xi_0) \widetilde{n_l}}{9\widetilde{n}}\,\left[\frac{1}{(3-\widetilde{n_l})^2} -
\frac{\widetilde{n_l}}{4\widetilde{T}}
\right]^{-1}\nonumber \\
&+~\frac{(\widetilde{n_g}-\widetilde{n_l})^2}{9\widetilde{n}}\,
\left[\frac{\widetilde{n_g} }{\xi_0\,(3- \widetilde{n_g})^2}-\frac{\widetilde{n_g}^2}{\xi_0\,4\widetilde{T}}
+ \frac{\widetilde{n_l} }{(1-\xi_0)(3-\widetilde{n_l})^2}-
\frac{\widetilde{n_l}^2}{(1-\xi_0)4\widetilde{T}}\right]^{-1}~ .
}

\begin{figure}[t]
\begin{minipage}{.8\textwidth}
\centering
\includegraphics[width=\textwidth]{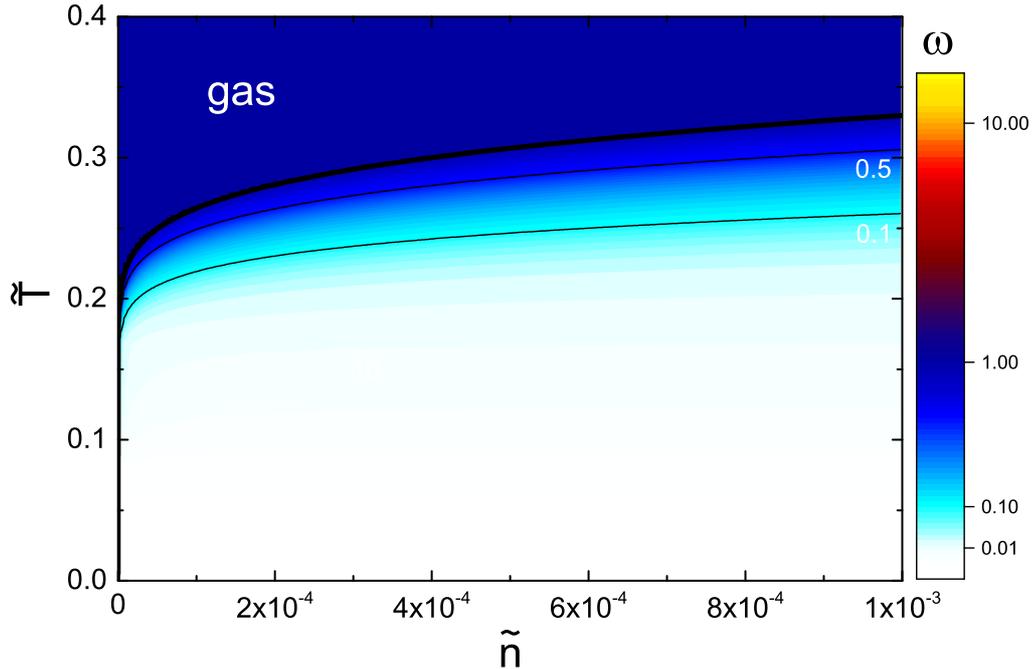}
\end{minipage}
\caption{(Color online) The lines of constant values of $\omega[N]$ are shown
at small temperatures and densities
on the $(\widetilde{n},\widetilde{T})$ phase diagram both inside and outside the mixed phase region.
 } \label{fig-small-n}
\end{figure}

In Fig.~\ref{fig-fluct-mp} the lines of constant values of $\omega[N]$ (\ref{omega-mix}) are shown
on the $(\widetilde{n},\widetilde{T})$ phase diagram inside the mixed phase.
At small $\widetilde{T}$, the particle number fluctuations inside the mixed phase are small,
$\omega[N]\ll 1$,
almost everywhere. This in because the main contribution to (\ref{omega-mix}) comes from the liquid component
of the mixed state, and the corresponding fluctuations inside the liquid phase are always small at small
$\widetilde{T}$. However, even at $\widetilde{T}\rightarrow 0$, the scaled variance
(\ref{omega-mix}) remains close to the ideal gas value, $\omega[N]\cong 1$,
in a small region near the boundary with the gaseous phase. This region is not seen
in Fig.~\ref{fig-fluct-mp} because of very small $\widetilde{n}$-values. A part of this region
with $0.2<\widetilde{T}<0.4$
(both inside and outside of the mixed phase) is shown in Fig.~\ref{fig-small-n}.

To study the behavior of (\ref{omega-mix}) in a vicinity of the critical point we introduce
$0<\rho_g=1-\widetilde{n_g}\ll 1$,~ $0<\rho_l=\widetilde{n_l}-1\ll 1$, and $0<t=1-\widetilde{T}\ll 1$.
At $t\rightarrow 0$ one finds
$\rho_g=\rho_l= 2\sqrt{t}$
%
%
and
\eq{\label{omegaNc}
\omega[N]~\cong~\frac{16}{9} ~t^{-1}~.
}
\begin{figure}[t]
\begin{minipage}{.98\textwidth}
\centering
\includegraphics[width=\textwidth]{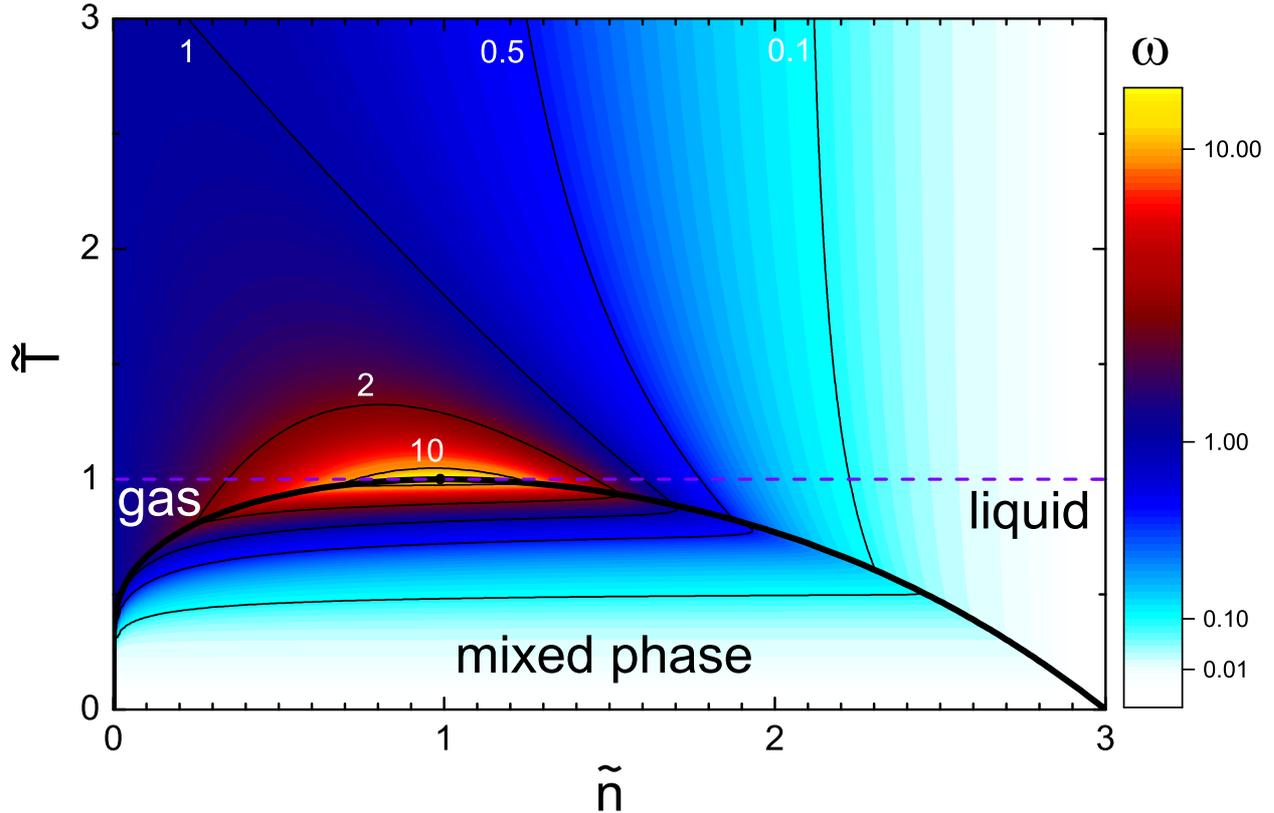}
\end{minipage}
\caption{(Color online)
The lines of constant values of $\omega[N]$ are shown
on the  phase diagram
for $0<\tilde{T}<3$
and for
all possible $\widetilde{n}$ values,
both inside and outside the mixed phase region.
 } \label{fig-fluct-both}
\end{figure}

Finally, the scaled variance $\omega[N]$ is shown in Fig.~\ref{fig-fluct-both}
for all possible values of $\widetilde{n}$ and $\widetilde{T}$, both inside and
outside of the mixed phase.
It is interesting to compare Eq.~(\ref{omega-mix}) for
the scaled variance calculated inside the mixed phase
and Eq.~(\ref{omega-2}) outside the mixed phase.
These two equations give the same results at the phase boundary.
At the boundary between the gaseous and mixed phase (i.e., $\xi_0=1$)
both the second and third terms in the right hand side of Eq.~(\ref{omega-mix})
become equal to zero, and the first term in (\ref{omega-mix})
coincides with Eq.~(\ref{omega-2}). Similarly,
at the boundary between the liquid and mixed phase (i.e., $\xi_0=0$)
both the first and third terms in the right hand side of Eq.~(\ref{omega-mix})
become equal to zero, and the second term in (\ref{omega-mix})
coincides with Eq.~(\ref{omega-2}).
It might seem from Fig.~\ref{fig-fluct-both} that at small temperatures fluctuations have a discontinuity at the boundary between the liquid and mixed phase.
It is, however, not the case: as mentioned above, when approaching the boundary from inside the mixed phase ($\xi_0 \to 0$), the third term in the r.h.s. of Eq.~(\ref{omega-mix}) vanishes, but at small temperatures it vanishes rapidly and only in the very vicinity of the phase boundary. Consequently, this behavior is seen poorly on the scale used in Fig.~\ref{fig-fluct-both}.
\section{Summary}\label{sec-sum}
The VDW pressure function (\ref{vdw-p-n}) has the
simple and familiar form in the CE.
However, there are some reasons for its reformulation in the GCE. First of all,
the CE pressure
does not give a complete thermodynamical
description of the system. Besides, the GCE has clear advantages if one will try to
extend the VDW model to multi-component hadron gas. The GCE formulation is also necessary
to find the particle number fluctuations in the VDW gas.

In the present paper we consider
the procedure of a transformation of the CE pressure
function to the GCE equation of state.
This procedure has a form of the first order differential equation for the free energy
$F$ as a function of $V$. A general solution of this equation includes the boundary condition
at $V=V_0$  which is taken in the form of
the ideal gas free energy in the limit of $N/V_0\rightarrow 0$.
This adds to the GCE formulation an information
of the particle masses and degeneracy
factors which was absent in the CE pressure function.
Following
this procedure we transform the CE VDW pressure (\ref{vdw-p-n}) to
the transcendental equation
(\ref{nvdw}) for the particle number density in the GCE.
Other thermodynamical functions of the VDW gas in the GCE can be then obtained in
a unique way.

The particle number fluctuations are calculated for the VDW gas in the GCE formulation.
An analytical expression for the scaled variance of particle number
fluctuations is derived in terms of dimensionless  particle number density $\widetilde{n}=n/n_c$
and temperature $\widetilde{T}=T/T_c$. It clearly demonstrates that in the VDW
gas the repulsive interactions suppress the particle number
fluctuations, whereas the attractive interactions lead to their
enhancement. In the pure phases far
away from
the critical point
the particle number fluctuations
approach to those of the ideal gas, i.e.,  $\omega[N]\cong 1$,
at $\widetilde{n}\ll 1$,
and becomes small, $\omega[N]\ll 1$, at $\widetilde{n}\rightarrow 3$.
The scaled variance $\omega[N]$ increases in a vicinity of the critical point and diverges
at the the critical point $\widetilde{n}=\widetilde{T}=1$.
Admitting a presence of metastable states, 
one observes that the scaled variance $\omega[N]$
diverges also at the boundary between the metastable and unstable regions, where
$\partial \widetilde{p} / \partial \widetilde{n} = 0$.

The fluctuations of partial volumes $V_g$ and $V_l$ occupied by the gaseous and liquid components
have
influence on the particle number fluctuations in the mixed phase.
However, the particle number  fluctuations remain finite in the mixed phase.  They start to increase
in a vicinity of the critical point and diverge at the critical point reached
from
inside the
mixed phase.

We hope that the present study of the VDW equation of state
within the GCE formulation
can be useful for several
applications in hadron physics. In particular, it looks promising
to consider the nuclear matter with
the VDW equation of state corrected by the Fermi statistics effects.
It is also interesting to study
the particle number fluctuations for hadronic systems
in a vicinity of the critical point.

\begin{acknowledgments}
We would like to thank
M. Ga\'zdzicki, S. Mr\'owczy\'nski, and K. Redlich for fruitful comments and discussions.
The work was partially supported
by HIC for FAIR within the LOEWE program of the State of Hesse
and by the Program of
Fundamental Research of the Department of Physics and Astronomy of NAS of Ukraine.
\end{acknowledgments}


\end{document}